\documentclass{nature}

\usepackage{graphicx}
\usepackage{dcolumn}
\usepackage[colorlinks=true,linkcolor=blue,citecolor=blue,urlcolor=blue]{hyperref}
\usepackage{stmaryrd}
\usepackage{esvect}
\usepackage{multirow}
\usepackage{xr}
\usepackage{upgreek}
\renewcommand{\vec}[1]{\mathbf{#1}}

\usepackage{siunitx}
\DeclareSIUnit\mub{\mu_\mathrm{B}}
\DeclareSIUnit\rydberg{\mathrm{Ry}}
\usepackage{mathtools}

\DeclarePairedDelimiterX\braket[1]{\langle}{\rangle}{#1}

\usepackage[usenames,dvipsnames]{xcolor}

\newcommand{\onlinecite}[1]{\hspace{-1 ex} \nocite{#1}\citenum{#1}}

\makeatletter
\let\saved@includegraphics\includegraphics
\AtBeginDocument{\let\includegraphics\saved@includegraphics}
\renewenvironment*{figure}{\@float{figure}}{\end@float}
\makeatother

\title{The chiral biquadratic pair interaction}
\author{Sascha Brinker$^{1,2}$, Manuel dos Santos Dias$^1$, Samir Lounis$^1$}

\begin{document}

\maketitle

\begin{affiliations}
 \item Peter Gr\"{u}nberg Institut and Institute for Advanced Simulation, Forschungszentrum J\"{u}lich \& JARA, 52425 J\"{u}lich, Germany
 \item Department of Physics, RWTH Aachen University, 52056 Aachen, Germany
\end{affiliations}


\begin{abstract}
\noindent
Magnetic interactions underpin a plethora of magnetic states of matter, hence playing a central role both in fundamental physics and for future spintronic and quantum computation devices.
The Dzyaloshinskii-Moriya interaction, being chiral and driven by relativistic effects, leads to the stabilization of highly-noncollinear spin textures such as skyrmions, which thanks to their topological nature are promising building blocks for magnetic data storage and processing elements.
Here, we reveal and study a new chiral pair interaction, which is the biquadratic equivalent of the Dzyaloshinskii-Moriya interaction.
First, we derive this interaction and its guiding principles from a microscopic model.
Second, we study its properties in the simplest prototypical systems, magnetic dimers deposited on various substrates, resorting to systematic first-principles calculations.
Lastly, we discuss its importance and implications not only for magnetic dimers but also for extended systems, namely one-dimensional spin spirals and complex two-dimensional magnetic structures, such as a nanoskyrmion lattice.
\end{abstract}

\newpage
\section*{Introduction}

\indent Starting from the seminal work of Heisenberg\cite{Heisenberg1928}, magnetic materials are often described by bilinear isotropic magnetic interactions, $J_{ij}\,\vec{S}_i\cdot\vec{S}_j$.
However, a wealth of complex spin-textures were discovered over the last century that called for the enrichment of the original Heisenberg model with various other types of interactions (see e.g. Refs.\cite{vanVleck1937,Dzyaloshinskii1958,Moriya1960,Anderson1963,Fawcett1988,Ruehrig1991,Katsura2005,Mostovoy2006,Lounis2008,Jackeli2009}). 
The magnetism of $^3$He is a striking example, being dominated by higher-order isotropic interactions\cite{Roger1983} which can be derived from the Hubbard model at half-filling\cite{Takahashi1977,MacDonald1988,Hoffmann2018} or from Kondo-lattice models\cite{Batista2016,Ozawa2017,Hayami2017}.  
These interactions, such as the biquadratic interaction $B_{ij}\,(\vec{S}_i\cdot\vec{S}_j)^2$ and the related three- and four-site interactions, introduce nonlinear effects into the Heisenberg model.
An important consequence is that different spin spirals, characterized by a wavevector $\vec{Q}$, can be combined into lower-energy multiple-$Q$-states, as the higher-order interactions invalidate the superposition principle.
Prominent examples are the antiferromagnetic $uudd$-state (a $2Q$-state)\cite{Al-Zubi2011,Kroenlein2018} and the $3Q$-state\cite{Kurz2001}.
Interestingly, this $3Q$-state (also magnetic skyrmions\cite{Bogdanov1989,Rossler2006} and bobbers\cite{Rybakov2015,Zheng2018}) is a noncoplanar magnetic state that hosts interesting Berry-phase physics arising from its non-vanishing scalar spin chirality $\vec{S}_i\cdot(\vec{S}_j\times\vec{S}_k)$, such as topological orbital ferromagnetism and Hall effects\cite{Taguchi2001,Shindou2001,Schulz2012,Dias2016,Hanke2017}.

The concept of vector spin chirality is embodied by the antisymmetric bilinear Dzyaloshinskii-Moriya interaction (DMI), $\vec{D}_{ij}\cdot(\vec{S}_i\times\vec{S}_j)$\cite{Dzyaloshinskii1958,Moriya1960}, which arises due to the combination of spin-orbit coupling and absence of spatial inversion symmetry.
The DMI lifts the energy degeneracy of magnetic spirals with opposite vector spin chirality, $\vec{S}_i\times\vec{S}_j$,  thus stabilizing magnetic structures of well-defined rotational sense, such as chiral spin spirals\cite{Bode2007,Ferriani2008} and magnetic skyrmions\cite{Bogdanov1989,Rossler2006}.
The intricate interplay of higher-order and anisotropic bilinear magnetic interactions generates various magnetic states: conical spin spirals\cite{Yoshida2012} and more complex magnetic structures\cite{Hanke2017,Takagi2018a,Romming2018}, such as an intricate nanoskyrmion lattice for a monolayer of Fe on the Ir(111) surface\cite{Heinze2011}.

In this work, we utilize a microscopic model combined with first-principles-based simulations to introduce and characterize a new kind of spin-orbit-driven magnetic pair interaction, the chiral biquadratic interaction (CBI).
It has the form $\vec{C}_{ij}\cdot(\vec{S}_i\times\vec{S}_j)\,(\vec{S}_i\cdot\vec{S}_j)$.
Like the DMI, this is a unidirectional interaction which is linear in the spin-orbit coupling, and so it is governed by the magnitude and orientation of the CBI vector $\vec{C}_{ij}$. We demonstrate that this vector obeys the same symmetry rules as the DMI\cite{Moriya1960,Fert1980,Levy1981,Crepieux1998}. 
Like the isotropic biquadratic interaction, it couples twice a pair of magnetic moments. After systematic investigations on magnetic dimers made of $3d$ elements on various surfaces with strong spin-orbit coupling, namely Pt(111), Pt(001), Ir(111) and Re(0001) surfaces, we find that the CBI can be comparable in magnitude to the DMI. Lastly, we explore the implications of the CBI for magnetic structures in one and two dimensions.

\section*{Results}

\subsection{Systematic microscopic derivation of higher-order interactions} \label{sec:systematic_derivation_higher_order_interactions}
The benefits of studying the properties of the magnetic interactions starting from a microscopic model are well-illustrated by the case of the DMI.
Although phenomenological arguments completely determine the form and symmetry properties of the DMI\cite{Dzyaloshinskii1958}, the microscopic analysis of Moriya\cite{Moriya1960} and later on the intuitive picture proposed by Fert and L\'evy\cite{Fert1980,Levy1981} have clarified the main ingredients that underpin this interaction.
We thus begin by introducing a generic model of the electronic structure of the magnetic material, and then outline how one can systematically extract all kinds of magnetic interactions from the electronic grand potential.

\noindent\emph{Microscopic model.} 
The microscopic hamiltonian that we consider has three contributions: $\mathcal{H} = \mathcal{H}^0 + \mathcal{H}^{\mathrm{mag}} + \mathcal{H}^{\mathrm{soc}}.$
Here $\mathcal{H}^0$ contains all spin-independent contributions, $\mathcal{H}^{\mathrm{mag}} = \sum_i U_i\,\vec{S}_i\cdot\vec{\upsigma}$ is the local exchange coupling of strength $U_i$ between the magnetic moment $\vec{S}_i$ on site $i$ and the electronic spin $\vec{\upsigma}$, and $\mathcal{H}^{\mathrm{soc}} = \sum_a \lambda_a\,\vec{L}_a\cdot\vec{\upsigma}$ is the atomic spin-orbit coupling of strength $\lambda_a$ on site $a$ between the electron spin and its atomic orbital angular momentum $\vec{L}_a$.
Grouping the spin-dependent terms into $\Delta\mathcal{H} = \mathcal{H}^{\mathrm{mag}} + \mathcal{H}^{\mathrm{soc}}$, it is straightforward to derive a formal power series for the electronic grand potential (see Supplementary Note 1),
    \begin{align}
        \Omega &= \Omega^0 - \frac{1}{\pi}\,\mathrm{Im}\!\int\!\mathrm{d} E \ f(E;\mu)
        \sum_p \frac{1}{p}\,\mathrm{Tr}\left[ \Delta \mathcal{H}\,G^0(E) \right]^p \nonumber\\
        &= \Omega^0 + \Omega^{\mathrm{soc}} + \sum_p \sum_{k=1}^{p/2} \Omega^{p,2k}[\{\vec{S}_i\}] \quad . \label{free_energy}
    \end{align}
    Here $\Omega^0$ is the contribution to the grand-canonical potential from the spin-independent $\mathcal{H}^0$, and $G^0(E) = (E - \mathcal{H}^0)^{-1}$ is the corresponding retarded Green function.
    The contributions arising solely from spin-orbit coupling are collected in $\Omega^{\mathrm{soc}}$, and the terms that depend on the magnetic moments are given by $\Omega^{p,2k}[\{\vec{S}_i\}]$.
    The Fermi-Dirac distribution for energy $E$ and chemical potential $\mu$ is given by $f(E;\mu)$, and the trace is over all sites, orbitals and spin degrees of freedom.

\noindent\emph{Diagrammatic rules.}
    \begin{figure}[!tb]
    \begin{center}
	\includegraphics[scale=1]{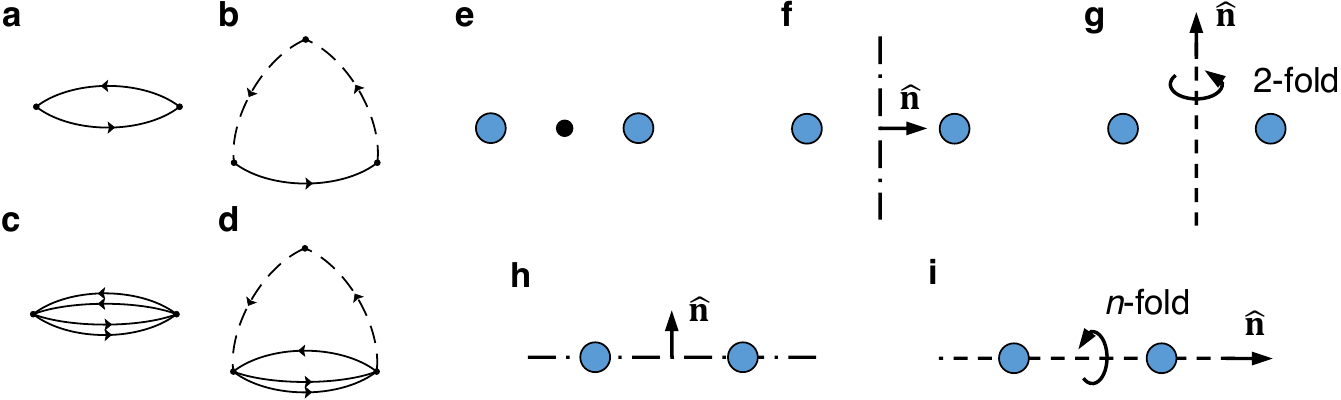}
	\caption{\label{fig:diagrams_and_symmetries_illustration} 	
	    Feynman diagrams and symmetry operations used for the microscopic derivation of the magnetic interactions.
	    (a-d) Prototypical diagrams for pair interactions up to fourth order in the magnetic hamiltonian and first order in the spin-orbit hamiltonian.
		Green functions connecting sites are represented by solid lines if both sites are magnetic and by dashed lines if one is a spin-orbit site.
		The five different symmetry operations being illustrated are:
		(e) inversion center in-between the two magnetic sites;
		(f) mirror plane perpendicular to their bond;
		(g) two-fold rotation perpendicular to their bond;
		(h) mirror plane containing both magnetic sites; and
		(i) $n$-fold rotation axis containing both magnetic sites.
		The unit vectors $\hat{\vec{n}}$ represent either the rotation axis or the normal to the mirror plane.
	}
    \end{center}
    \end{figure}
    Only a subset of the terms contained in $\Omega^{p,2k}[\{\vec{S}_i\}]$ are of interest for the purpose of identifying the possible types of magnetic interactions.
    As detailed in Supplementary Note 1, these can be represented by prototypical diagrams for all kinds of magnetic interactions, from magnetocrystalline anisotropies to pair interactions or many-site interactions.
    Each diagram contain $p$ vertices connected by $p$ lines.
    The vertices in a prototypical diagram must correspond to spatially distinct sites, and the lines represent connections between the sites through $G^0$.
    Each diagram contains $2k$ magnetic sites and $p-2k$ spin-orbit sites.
    A link between two magnetic sites is denoted by a solid line, while a link between a magnetic site and a spin-orbit site is marked by a dashed line.
    A magnetic site cannot appear consecutively (i.e.\ a line cannot close on itself), and two consecutive spin-orbit sites (distinct or not) are also excluded.
    In this work we focus on interactions involving two magnetic sites, and the corresponding prototypical diagrams up to fourth order in the magnetic sites and first order in spin-orbit coupling are shown in Fig.~\ref{fig:diagrams_and_symmetries_illustration}a-d.

\noindent\emph{Prototypical diagrams.}
    It is a simple matter to extract the form of the magnetic interactions from each prototypical diagram, by using the properties of traces of Pauli matrices.
    The derivations and the forms of the coupling coefficients can be found in Supplementary Note 1.
    The first and simplest diagram is given in Fig.~\ref{fig:diagrams_and_symmetries_illustration}a and translates into the isotropic bilinear exchange interaction $\frac{1}{2} \sum_{i,j} J_{ij}\,\vec{S}_i \cdot \vec{S}_j$.
    Attaching one spin-orbit site to the diagram of Fig.~\ref{fig:diagrams_and_symmetries_illustration}a results in the diagram shown in Fig.~\ref{fig:diagrams_and_symmetries_illustration}b.
    This generates the DMI, $\frac{1}{2} \sum_{i,j}  \vec{D}_{ij} \cdot \left( \vec{S}_i \times \vec{S}_j \right)$.
    The structure of this diagram is identical to the third-order perturbation theory developed by Fert and L\'evy\cite{Fert1980,Levy1981}.
    The DMI vector is determined by the properties and geometrical arrangement of the spin-orbit sites, $\vec{D}_{ij} = \sum_a \vec{D}_{ij,a}$.
    Due to the cross product form, it favors magnetic structures with a definite vector chirality.
    The next diagram is shown in Fig.~\ref{fig:diagrams_and_symmetries_illustration}c and leads to the isotropic biquadratic interaction, $\frac{1}{2} \sum_{i,j} B_{ij} \left(\vec{S}_i \cdot \vec{S}_j\right)^2$.
    The diagrams with the same number of lines but connecting either three or four different magnetic sites lead to the isotropic 4-spin 3-site and 4-spin 4-site interactions, respectively (see Supplementary Note 1).
    Lastly, we find a new kind of magnetic interaction from the prototypical diagram shown in Fig.~\ref{fig:diagrams_and_symmetries_illustration}d:
    \begin{align}
        \mathrm{(1d)} \;\rightarrow\; \frac{1}{2} \sum_{i,j} 
        \vec{C}_{ij} \cdot \left( \vec{S}_i \times \vec{S}_j  \right)  \left( \vec{S}_i \cdot \vec{S}_j \right)
        \quad . \label{diagram_biquadratic_DMI}
    \end{align}
    We name it the chiral biquadratic interaction (CBI), as it is an antisymmetric 4-spin 2-site interaction generated by an additional spin-orbit site.
    It thus combines the isotropic scalar product $\vec{S}_i \cdot \vec{S}_j$ with the chiral coupling $\vec{C}_{ij}\cdot\left( \vec{S}_i \times \vec{S}_j \right)$ defined by the CBI vector $\vec{C}_{ij} = \sum_a \vec{C}_{ij,a}$, which is generated by the spin-orbit sites.
    This is our main quantity of interest and its properties will be discussed in detail in this paper.
    The diagrams with the same number of lines but connecting either three or four different magnetic sites lead to the chiral 4-spin 3-site and 4-spin 4-site interactions, respectively (see Supplementary Note 1).
    
\noindent\emph{Symmetry rules.} 
    We next study what are the properties of the newly-found CBI vector, $\vec{C}_{ij}$, by comparison with those of the DMI vector, $\vec{D}_{ij}$.
    Given a pair of magnetic sites $i$ and $j$ connected with the vector $\vec{R}_{ij}$, there are five relevant symmetries, which are illustrated in Fig.~\ref{fig:diagrams_and_symmetries_illustration}e-i.
    Within the picture of the prototypical diagrams, these symmetry operations are a combination of local transformations at each site (e.g.\ a rotation or a mirroring) and a permutation of the sites.
    Importantly, symmetry dictates what is the spatial arrangement of the spin-orbit sites around the pair of magnetic sites.
    One can then relate the diagrams connecting the pair of magnetic sites to each spin-orbit site, noting that the orbital angular momentum operator transforms as a pseudovector, and from this derive the symmetry rules for each magnetic interaction.
    For the DMI vector, these symmetries lead to the so-called Moriya's rules\cite{Moriya1960,Levy1981,Crepieux1998}.
    These rules are, for each symmetry operation shown in Fig.~\ref{fig:diagrams_and_symmetries_illustration}e-i: (e) $\vec{D}_{ij}=0$, (f) $\mathcal{P}_{\hat{\vec{n}}}^\parallel\,\vec{D}_{ij} = 0$, (g) $\mathcal{P}_{\hat{\vec{n}}}^\parallel\,\vec{D}_{ij} = 0$, (h) $\mathcal{P}_{\hat{\vec{n}}}^\perp \vec{D}_{ij} = 0$, and (i) $\mathcal{P}_{\hat{\vec{n}}}^\perp \vec{D}_{ij} = 0$  (see Supplementary Note 2).
    The vanishing components of the DMI vector are those either parallel or perpendicular to $\hat{\vec{n}}$, which represents either the rotation axis or the normal to the mirror plane.
    It follows naturally from comparing the structure of the prototypical diagrams for the DMI and the CBI that precisely the same rules apply to the CBI vector, $\vec{C}_{ij}$.
    The two vectors do not have to be collinear, notably if the only applicable symmetry is of type (f).

\noindent\emph{Connection to a phenomenological model.}
Another advantage of of our approach is apparent if we consider the appropriate phenomenological model for the magnetic interactions.
To illustrate this point, we consider the most general spin model containing only bilinear and biquadratic pair interactions:
\begin{equation}\label{eq:hamil_pair}
    \mathcal{H}^{\mathrm{pair}} = \frac{1}{2} \sum_{i,j}\sum_{\alpha,\beta} J_{ij}^{\alpha\beta} S_i^\alpha S_j^\beta
    + \frac{1}{2} \sum_{i,j}\sum_{\alpha,\beta,\gamma,\delta} B_{ij}^{\alpha\beta\gamma\delta} S_i^\alpha S_j^\beta S_i^\gamma S_j^\delta \quad .
\end{equation}
The bilinear interactions are described by a rank-2 cartesian tensor $J_{ij}^{\alpha\beta}$ (9 parameters), which contains the isotropic pair interaction given by $J_{ij}\,\vec{S}_i \cdot \vec{S}_j$ (1 parameter), the DMI given by $\vec{D}_{ij} \cdot \left( \vec{S}_i \times \vec{S}_j \right)$ (3 parameters), and the remaining five parameters describe the symmetric bilinear pair anisotropy.
The biquadratic interactions are described by a rank-4 cartesian tensor $B_{ij}^{\alpha\beta\gamma\delta}$ (81 parameters), and are not straightforward to classify.
The number of independent elements of the biquadratic tensor is reduced to 25 by noting that $B_{ij}^{\alpha \beta \gamma \delta} = B_{ij}^{\gamma \beta \alpha \delta} = B_{ij}^{\alpha  \delta \gamma \beta} = B_{ij}^{ \gamma \delta \alpha \beta} $
and that excluding terms which are independent of the spin orientation requires $\sum_{\alpha} B_{ij}^{\alpha \beta \alpha \delta} = \sum_{\beta}  B_{ij}^{\alpha \beta \gamma \beta} = 0$.
The same conclusion as to the number of independent parameters can be arrived at via the spin cluster expansion of the magnetic energy\cite{Drautz2004,Drautz2005} (see Supplementary Note 3).
Making use of the prototypical diagrams, we already recovered the isotropic biquadratic interaction $B_{ij} \left(\vec{S}_i \cdot \vec{S}_j\right)^2$ (1 parameter), and we uncovered the CBI given by $\vec{C}_{ij} \cdot \left( \vec{S}_i \times \vec{S}_j  \right)  \left( \vec{S}_i \cdot \vec{S}_j \right)$ (3 parameters).
Considering prototypical diagrams with more spin-orbit sites is a constructive approach to populate the rest of the $B_{ij}^{\alpha \beta \gamma \delta}$ tensor, from which the form of the magnetic interactions will also follow.
If spin-orbit coupling is weak in some sense, we then also obtain a natural classification of the various magnetic interactions in powers of this small parameter.
This would justify considering only the isotropic biquadratic and the CBI as the most important interactions among all biquadratic ones.
Instead, we shall turn to realistic calculations of the magnitude and properties of the magnetic interactions.


\subsection{Magnetic interactions of dimers on various surfaces}

In order to quantify the properties and significance of the CBI in relation to the other magnetic interactions, we present a systematic study of a series of prototypical systems: magnetic dimers on several surfaces for which the spin-orbit effects are strong.
To do so, we construct a complete magnetic model containing all relevant interactions up to four-spin couplings by defining a mapping from a set of self-consistent constrained DFT calculations, as explained in the Methods and Supplementary Note 4.
The parametrizations of the complete magnetic model for all considered systems are given in Supplementary Note 5.

\begin{figure}[!t]
    \begin{center}
	\includegraphics[scale=0.9]{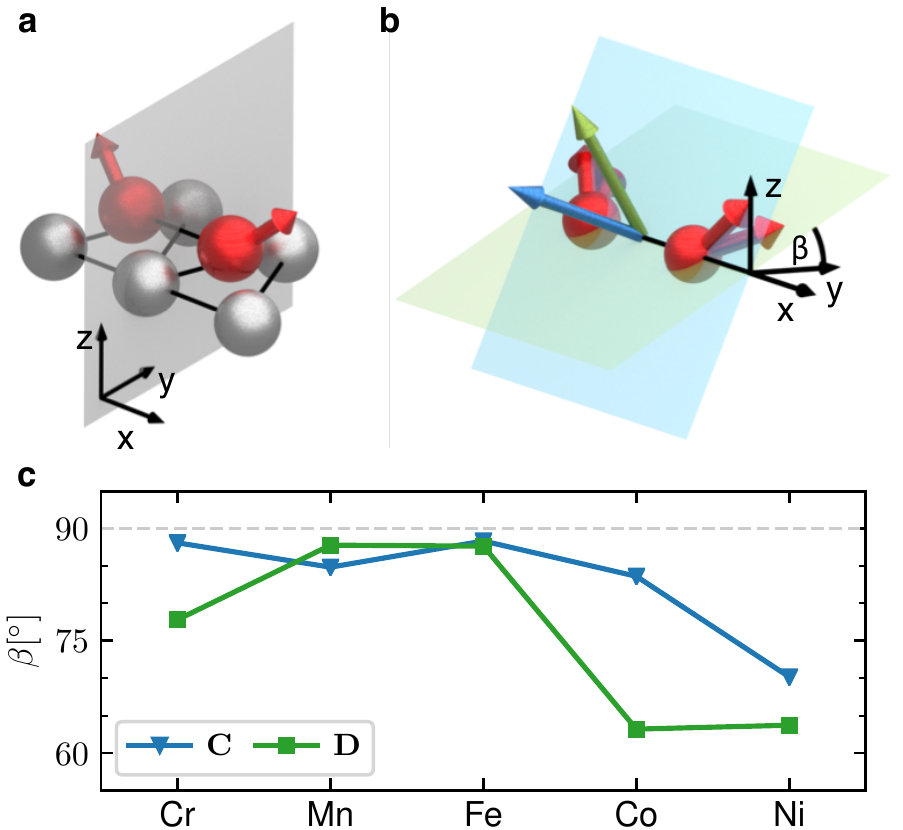}
	\caption{\label{fig:angle_DMI} 
	    Illustration of the chiral interactions in a magnetic dimer.
	    (a) Geometry: the dimer axis is along the $x$-direction, the surface normal defines the $z$-axis, and the mirror plane (grey) perpendicular to the bond is also shown.
	    The atoms are represented by red spheres and their spin magnetic moments by solid arrows.
	    Neighboring surface atoms are represented by grey spheres.
		(b) The CBI (blue) and DMI (green) vectors are shown as possibly noncollinear, and the dimer magnetic structures that they favor is also indicated.
		The plane containing the magnetic structure is characterized by its dihedral angle $\beta$ with the $xy$-plane.
		(c) Dihedral angle $\beta$ between the plane generated by the CBI or DMI vectors and the $xy$-plane for the Cr, Mn, Fe, Co and Ni dimers deposited on Pt(111).
	}
    \end{center}
\end{figure}

\noindent\emph{Simplified magnetic model.}
We focus on the following interactions: the isotropic bilinear interaction $J\,\hat{\vec{e}}_1\cdot\hat{\vec{e}}_2$, the DMI $\vec{D}\cdot(\hat{\vec{e}}_1\times\hat{\vec{e}}_2)$, the isotropic biquadratic interaction $B\,(\hat{\vec{e}}_1\cdot\hat{\vec{e}}_2)^2$, and the CBI $\vec{C}\cdot(\hat{\vec{e}}_1\times\hat{\vec{e}}_2)\,(\hat{\vec{e}}_1\cdot\hat{\vec{e}}_2)$.
These are defined in terms of the unit vectors $\hat{\vec{e}}_i$ representing the orientation of the spin magnetic moment of the $i$-th atom in the dimer.
The axis of the dimer is chosen as the $x$-axis, while the normal to the surface is chosen as the $z$-axis, as shown in Fig.~\ref{fig:angle_DMI}a.
Symmetry then restricts the DMI and CBI vectors to lie in the $yz$-plane for (111) and (0001) surfaces, illustrated in in Fig.~\ref{fig:angle_DMI}b, or to lie along the $y$-axis for the (001) surface.
In all cases the $y$-component is the dominant one.
To understand the interplay between the different interactions, we consider the simplified model obtained from Eq.~\eqref{eq:hamil_pair} by confining the magnetic moments to the $xz$-plane and keeping only the mentioned interactions:
\begin{align}\label{eq:energy}
    E(\alpha) = J\cos\alpha + D_y\sin\alpha + B\cos^2\alpha + C_y\sin\alpha\cos\alpha \quad.
\end{align}
Here $\alpha = \theta_2 - \theta_1$ is the opening angle between the two magnetic moments.
The angle that minimizes the energy can be written as $\alpha_{\mathrm{min}} = \alpha_J + \Delta\alpha$, where $\alpha_J = \SI{0}{\degree}$ if $J<0$ (ferromagnetic) or $\SI{180}{\degree}$ if $J>0$ (antiferromagnetic), and $\Delta\alpha$ is the canting induced by the remaining magnetic interactions.
The energy can then be expanded as
\begin{align}\label{eq:canting}
    E(\alpha_{\mathrm{min}}) &\approx E(\alpha_J) - \big(\mathrm{sgn}(J) D_y - C_y\big)\,\Delta\alpha + \big(|J| - 2B\big)\,\frac{(\Delta\alpha)^2}{2} \\ 
    &\Rightarrow \Delta\alpha = \frac{\SI{180}{\degree}}{\pi}\,\frac{\mathrm{sgn}(J) D_y - C_y}{|J| - 2B} \quad,
\end{align}
where the last line gives an approximation to the canting angle.

\begin{table}
    \renewcommand{\arraystretch}{0.7}
    \centering
    \begin{tabular}{rc|ccccccccc}
        \multicolumn{2}{c|}{Dimer} & $M$  ($\mu_{\mathrm{B}}$) & $C_y$ & $D_y$ & $B$ & $J$ &
        $\Delta\alpha^{\mathrm{2s}}$ & $\Delta\alpha^{\mathrm{4s}}$ & $D^{\mathrm{eff}}_y$ & $J^\mathrm{eff}$ \\ \hline
        \multirow{5}{*}{Pt(111)} 
        & Cr & $3.26$ & $\phantom{-}\SI{2.5}{}$ & $\phantom{-0}\SI{8.5}{}$ & $\SI{-11.7}{}$ & $\phantom{-}\SI{35.8}{}$ &
        $\SI{-13}{\degree}$ & $\phantom{0}\SI{-6}{\degree}$ & $\phantom{-0}\SI{6.0}{}$ & $\phantom{-}\SI{59.2}{}$ \\
        & Mn & $4.05$ & $\SI{-0.6}{}$ & $\phantom{0}\SI{-3.3}{}$ & $\phantom{-0}\SI{0.8}{}$ & $\phantom{-}\SI{58.5}{}$ &
        $\phantom{-0}\SI{3}{\degree}$ & $\phantom{-0}\SI{3}{\degree}$ & $\phantom{0}\SI{-2.7}{}$ & $\phantom{-}\SI{56.9}{}$ \\
        & Fe & $3.32$ & $\phantom{-}\SI{2.6}{}$ & $\phantom{0}\SI{-7.3}{}$ & $\phantom{0}\SI{-2.2}{}$ & $\SI{-43.0}{}$ &
        $\phantom{-}\SI{10}{\degree}$ & $\phantom{-0}\SI{6}{\degree}$ & $\phantom{0}\SI{-4.7}{}$ & $\SI{-47.4}{}$ \\
        & Co & $2.12$ & $\SI{-1.6}{}$ & $\phantom{-0}\SI{7.1}{}$ & $\phantom{-0}\SI{0.8}{}$ & $\SI{-76.8}{}$ &
        $\phantom{0}\SI{-5}{\degree}$ & $\phantom{0}\SI{-4}{\degree}$ & $\phantom{-0}\SI{5.5}{}$ & $\SI{-75.2}{}$ \\
        & Ni & $0.62$ & $\phantom{-}\SI{0.5}{}$ & $\phantom{-0}\SI{0.8}{}$ & $\phantom{0}\SI{-1.1}{}$ & $\phantom{0}\SI{-5.4}{}$ &
        $\phantom{0}\SI{-8}{\degree}$ & $\SI{-10}{\degree}$ & $\phantom{-0}\SI{1.3}{}$ & $\phantom{0}\SI{-7.6}{}$ \\ \hline
        \multirow{2}{*}{Pt(001)}  
        & Cr & $2.53$ & $\phantom{-}\SI{2.5}{}$ & $\phantom{-}\SI{11.2}{}$ & $\phantom{0}\SI{-9.7}{}$ & $\SI{-35.3}{}$ &
        $\phantom{}\SI{-18}{\degree}$ & $\phantom{}\SI{-14}{\degree}$ & $\phantom{-}\SI{13.7}{}$ & $\SI{-54.7}{}$\\ 
        & Fe & $3.24$ & $\SI{-0.2}{}$ & $\phantom{0}\SI{-9.5}{}$ & $\phantom{0}\SI{-1.5}{}$ & $\phantom{-}\SI{15.0}{}$ &
        $\phantom{-}\SI{32}{\degree}$ & $\phantom{-}\SI{28}{\degree}$ & $\phantom{0}\SI{-9.3}{}$ & $\phantom{-}\SI{12.0}{}$\\ \hline
        \multirow{2}{*}{Ir(111)} 
        & Cr & $3.02$ & $\phantom{-}\SI{3.2}{}$ & $\phantom{-}\SI{10.7}{}$ & $\SI{-12.1}{}$ & $\phantom{-}\SI{29.5}{}$ &
        $\phantom{}\SI{-20}{\degree}$ & $\phantom{0}\SI{-8}{\degree}$ & $\phantom{-0}\SI{7.5}{}$ & $\phantom{-}\SI{53.7}{}$\\
        & Fe & $3.06$ & $\phantom{-}\SI{1.3}{}$ & $\SI{-14.6}{}$ & $\phantom{0}\SI{-3.6}{}$ & $\SI{-16.3}{}$ &
        $\phantom{-}\SI{42}{\degree}$ & $\phantom{-}\SI{32}{\degree}$ & $\phantom{}\SI{-13.3}{}$ & $\phantom{}\SI{-23.3}{}$\\ \hline
        \multirow{2}{*}{Re(0001)} 
        & Cr & $2.18$ & $\phantom{-}\SI{0.4}{}$ & $\SI{-18.1}{}$ & $\phantom{0}\SI{-3.4}{}$ & $\SI{-16.4}{}$ &
        $\phantom{-}\SI{48}{\degree}$ & $\phantom{-}\SI{40}{\degree}$ & $\phantom{}\SI{-17.7}{}$ & $\phantom{}\SI{-23.2}{}$\\
        & Fe & $2.29$ & $\phantom{-}\SI{0.3}{}$ & $\phantom{-0}\SI{0.5}{}$ & $\phantom{-0}\SI{0.1}{}$ & $\phantom{0}\SI{-2.3}{}$ &
        $\phantom{}\SI{-12}{\degree}$ & $\phantom{}\SI{-19}{\degree}$ & $\phantom{-0}\SI{0.8}{}$ & $\phantom{0}\SI{-2.5}{}$ \\
    \end{tabular}
	\caption{Spin moments, magnetic interaction parameters and opening angles of the magnetic ground state for Cr, Mn, Fe, Co and Ni dimers deposited on Pt(111).
	$M$ is the spin magnetic moment of one atom in the dimer.
	The CBI and the DMI are represented by their dominant vector component $C_y$ and $D_y$, respectively.
	The biquadratic and bilinear isotropic interactions are given by $B$ and $J$, respectively.
	All interaction values are in meV.
	The canting angles are found by minimizing Eq.~\eqref{eq:energy} with all the interactions ($\Delta\alpha^{\mathrm{4s}}$) or keeping only $J$ and $D_y$ ($\Delta\alpha^{\mathrm{2s}}$).
	The sign of $\Delta\alpha$ represents the sign of $(\hat{\vec{e}}_1 \times \hat{\vec{e}}_2)_y$, the vector chirality of the magnetic ground state.
	For comparison, we also list the effective bilinear interactions defined by $D_y^{\mathrm{eff}} = D_y  - \mathrm{sgn}(J) C_y$ and $J^{\mathrm{eff}} = J - 2\,\mathrm{sgn}(J) B$.
	}
	\label{tab:dimer_exchange_parameters}
\end{table}

\noindent\emph{Magnetic dimers on Pt(111).}
We first compare the magnetic properties of five different homoatomic dimers on the Pt(111) surface, with the corresponding data collected in Table~\ref{tab:dimer_exchange_parameters}.
All dimers except Ni possess large spin magnetic moments, which depend very weakly on the various imposed magnetic structures.
Comparing the CBI to the DMI, we see that the magnitude of $C_y$ is around $20$--$30\%$ of the one of $D_y$, even reaching $60\%$ for Ni.
For most dimers, $B$ is similar in magnitude to the CBI, and is even stronger than the DMI for Cr and Ni.
According to $J$, which is the dominant interaction, Cr and Mn are antiferromagnetic, while Fe, Co and Ni are ferromagnetic.
Considering only $J$ and $D_y$ leads to a canting of the magnetic structure given by $\Delta\alpha^{\mathrm{2s}}$ in Table~\ref{tab:dimer_exchange_parameters}, while considering also $B$ and $C_y$ we obtain $\Delta\alpha^{\mathrm{4s}}$.
The difference between these values is the largest for Cr and Fe, so these are the dimers for which the biquadratic interactions are most important.
Lastly, we also include the values of the effective bilinear interactions defined by the coefficients of $\Delta\alpha$ and $(\Delta\alpha)^2/2$ in Eq.~\eqref{eq:canting}.
These correspond to $J^{\mathrm{eff}} = J - 2\,\mathrm{sgn}(J) B$ and $D_y^{\mathrm{eff}} = D_y  - \mathrm{sgn}(J) C_y$.
The vector chirality of the magnetic ground state is set by the combination of the DMI and CBI vectors.
These can be parallel, antiparallel, or substantially noncollinear (shown in Fig.~\ref{fig:angle_DMI}c), in particular for the Co dimer.
This shows that the CBI has not only the potential to impose the opposite vector spin chirality to the one favoured by the DMI ($\Delta\alpha$ changing sign in Eq.~\eqref{eq:canting}), but also to tilt in away from the direction defined by the DMI vector.

\noindent\emph{Cr and Fe dimers on other surfaces.}
The Cr and Fe dimers on Pt(111) were found to have the most important contributions from the CBI.
To ascertain whether this is particular to the Pt(111) surface, we placed these dimers on other surfaces with strong spin-orbit coupling, namely Pt(001), Ir(111) and Re(0001).
We see from Table~\ref{tab:dimer_exchange_parameters} that the CBI is generally a sizeable fraction of the DMI.
On the Pt(001) and Ir(111) surfaces, the two dimers display a very large DMI, even in relation to its isotropic bilinear interaction $J$, leading to a strong canting of the magnetic structure.
This canting is substantially modified when the biquadratic interactions are accounted for.
The same behavior is found for the Cr dimer on Re(0001), while for the Fe dimer on this surface the interactions are found to be surprisingly weak, but still support a strongly noncollinear magnetic structure.

\begin{figure}[!t]
    \begin{center}
	\includegraphics[scale=0.97]{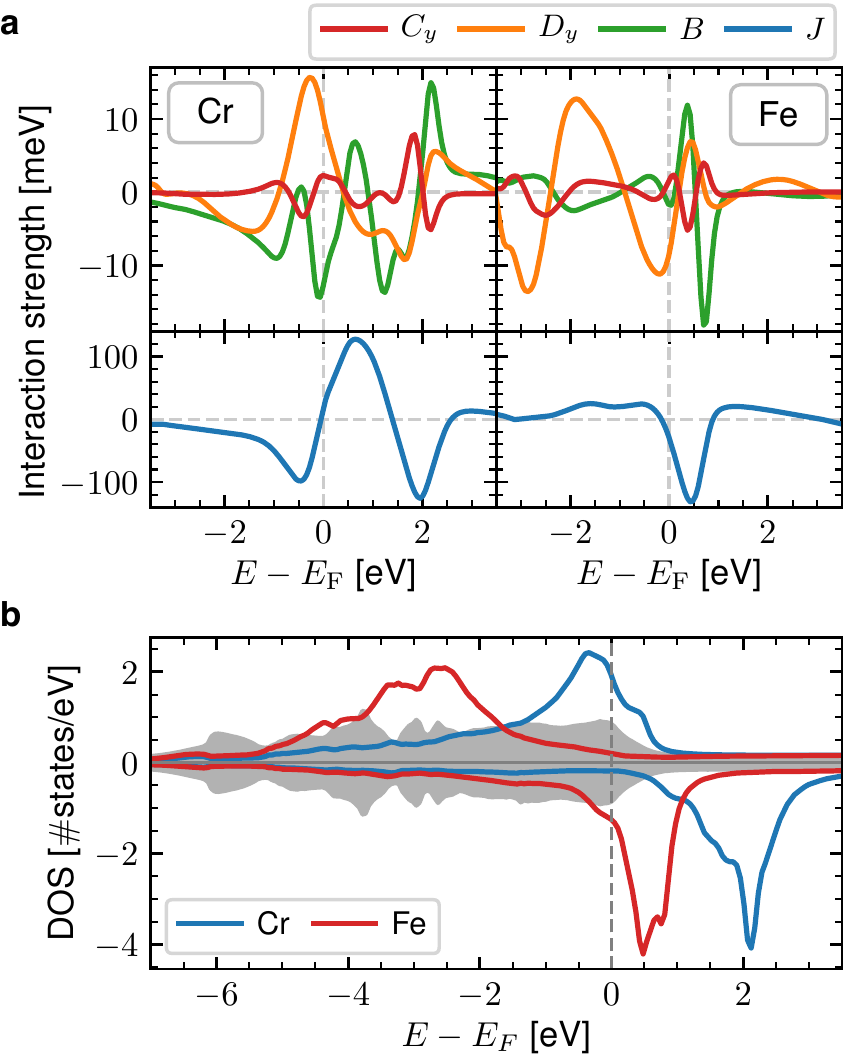}
	\caption{\label{fig:energy_dep_DMI_noconstr} 
	    Relation between the magnetic interactions and the electronic structure of Cr and Fe dimers on Pt(111).
		(a) Energy dependence of the magnetic interactions in an energy window around the Fermi level for a Cr and an Fe dimer deposited on the Pt(111) surface.
	   	(b) Local density of states for an atom of the Cr and Fe dimers deposited on the Pt(111) surface.
		The local density of states of the Pt surface is shown as a grey background.
		The majority spin contribution is shown as positive and the minority as negative values.
	}
    \end{center}
\end{figure}

\noindent\emph{Electronic origin of the magnetic interactions.}
The origin of the different magnetic interactions can be further understood by comparing their dependence on the filling of the electronic states with the corresponding density of states of each dimer.
This is shown in Fig.~\ref{fig:energy_dep_DMI_noconstr} for the Cr and Fe dimers on Pt(111).
$D_y$ is largest in the energy range of the Pt $d$-states, which shows that the DMI is strongly enhanced by hybridization of the magnetic $d$-states of the dimer with the $d$-states of the Pt surface.
$B$ and $C_y$ (and also $J$) have their largest values in the small energy range of the minority $d$-states of the Fe dimer, which suggests that the biquadratic interactions require less hybridization with Pt and so more localized magnetic $d$-states in the dimer.
This can be understood from the microscopic theory.
According to the prototypical diagrams in Fig.~\ref{fig:diagrams_and_symmetries_illustration}a,c, the isotropic interactions are direct interactions between the dimer atoms, resulting in a strong dependence on the Fe $d$-states.
The prototypical diagram of Fig.~\ref{fig:diagrams_and_symmetries_illustration}b shows that the DMI is an interaction mediated by a spin-orbit site, which are supplied by the Pt surface atoms, and so this interaction is strongly dependent on the Pt $d$-states.
The CBI involves both a direct exchange between the magnetic sites and an excursion through a spin-orbit site, Fig.~\ref{fig:diagrams_and_symmetries_illustration}d, so it can be amplified in those two ways, leading to a more complicated dependence.


\subsection{Implications of the chiral biquadratic interaction}
The CBI has different important implications for a broad class of noncollinear magnetic nanostructures.
For a magnetic dimer, we already found that the CBI influences the opening angle and the vector spin chirality of the magnetic structure.
To gain further understanding, we return to the previous example of a CBI vector in the $y$-direction, for which the interaction energy has the form $E_{\mathrm{CBI}}(\alpha) = C_y\sin\alpha\cos\alpha$ (see Eq.~\eqref{eq:energy}).
The $\cos\alpha$ term comes from the $\hat{\vec{e}}_1\cdot\hat{\vec{e}}_2$ part of the interaction, while the $\sin\alpha$ terms comes from $(\hat{\vec{e}}_1\times\hat{\vec{e}}_2)_y$, with $\alpha$ the opening angle.
As the dot product is isotropic, the opening is favored in the plane perpendicular to the CBI vector.
Fixing $C_y > 0$ for definiteness, there are two energy minima for $\alpha_{\mathrm{min}} \in \{\SI{-45}{\degree}, \SI{135}{\degree}\}$, and two maxima for $\alpha_{\mathrm{max}} \in \{\SI{45}{\degree}, \SI{-135}{\degree}\}$.
Strikingly, the two values of $\alpha_{\mathrm{min}}$ have opposite signs, which means that the sign of their vector spin chirality (projected on the $y$-axis) is also opposite.
Thus, and in contrast to the DMI, the CBI favors both possible rotational senses at once (although with different opening angles).
Starting from a ferromagnetic or antiferromagnetic structure (set by $J$), the DMI will induce a canting of the same rotational sense for both cases, while the CBI will favor cantings for each structure which have opposite rotational senses.

\begin{figure}[!t]
    \begin{center}
	\includegraphics[scale=1]{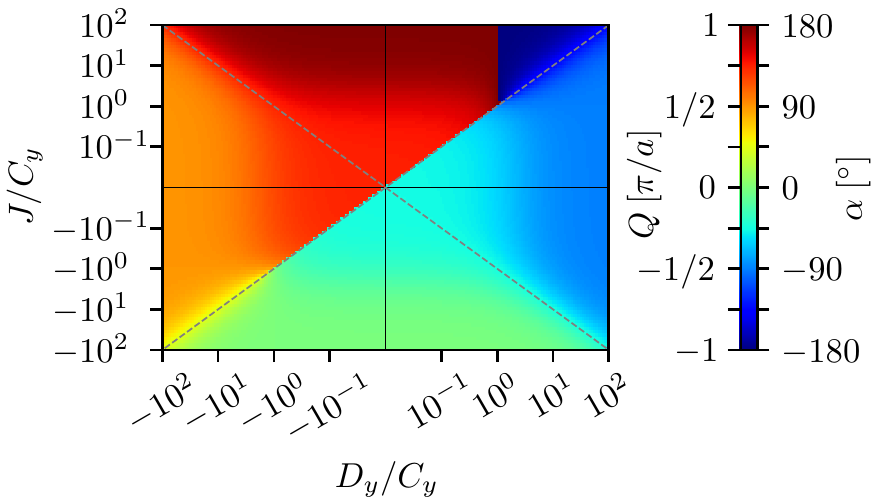}
	\caption{\label{fig:BDMI_impact}
	    Magnetic ground states of a dimer and of an infinite chain stabilized by the isotropic bilinear interaction $J$, the DMI $D_y$ and the CBI $C_y$.
	    The magnetic ground state is characterized by the opening angle $\alpha$ between two neighboring magnetic moments, obtained by minimizing Eq.~\eqref{eq:energy}, which for an infinite chain defines a spin spiral with wavevector $Q = \alpha/a$ ($a$ being the nearest-neighbor distance).
	    $C_y$ is chosen to be positive.
	}
    \end{center}
\end{figure}

\noindent\emph{From a dimer to an infinite chain.}
Next we relate the magnetic ground state of a dimer to that of an infinite chain, assuming that the interactions present are $J$, $D_y$ and $C_y$ (we take $B=0$ for simplicity), being nearest-neighbor interactions for the chain.
The energy as a function of the opening angle for the dimer is given by Eq.~\eqref{eq:energy}, and the same form applies for the energy of a spiral magnetic structure on an infinite chain, with $\vec{\hat{e}}_i = \sin(Qx_i)\,\vec{\hat{e}}_x+\cos(Qx_i)\,\vec{\hat{e}}_z$, by setting $Q = \alpha/a$ ($a$ is the nearest-neighbor distance).
The canting angle/spiral wavevector that minimizes the energy is shown for different values of $D_y/C_y$ and $J/C_y$ in Fig.~\ref{fig:BDMI_impact}.
The sharp transition in the sign of $\alpha$ found along $J=D_y$ is driven by $C_y$, and signals a change in the sign of the vector spin chirality. 
At the edges of the diagram we recover well-known magnetic structures ($\alpha=\SI{0}{\degree},\SI{180}{\degree}$ are the ferromagnetic and antiferromagnetic structures for $J$-only, respectively, and $\alpha=\pm\SI{90}{\degree}$ pertain to $D_y$-only).

\begin{figure}[!t]
    \begin{center}
	\includegraphics[scale=0.97]{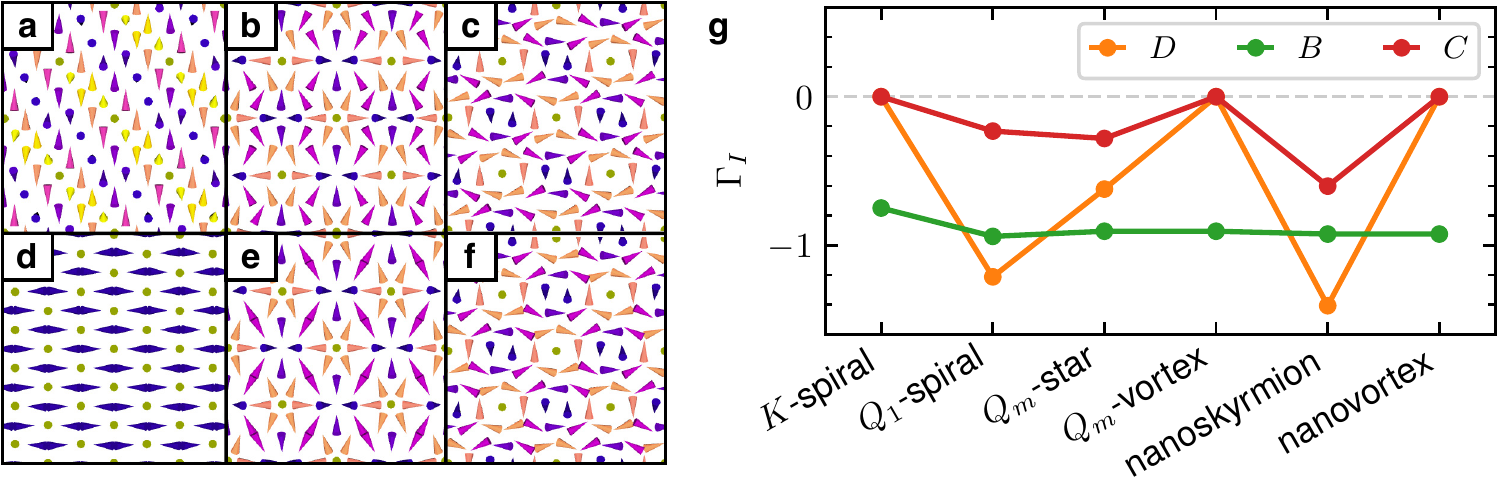}
	\caption{\label{fig:energy_dependence_structures}
	Contributions from different magnetic interactions to the energy of various magnetic structures.
	The considered magnetic structures are: single-$Q$ spin spirals for the (a) $Q_1$-vector and (d) and for the $K$-vector (N\'eel state); the $2Q$-states (b) $Q_m$-star and (c) $Q_m$-vortex; and the two noncollinear states forming (e) a nanoskyrmion lattice and (f) a nanovortex lattice.
	(g) The $\Gamma_I$ factors describe how the contribution to the energy from different interactions varies among the several magnetic structures (see main text for discussion).
	The considered interactions are the DMI $D$, the isotropic biquadratic interaction $B$ and the CBI $C$.
	These magnetic structures were discussed in connection to the magnetic ground state of an Fe monolayer on Ir(111) in Ref.~\onlinecite{Heinze2011}.
	}
    \end{center}
\end{figure}

\noindent\emph{Impact of the CBI on complex 2D magnetic structures.}
As a final example, we consider two-dimensional magnetic systems.
Higher-order isotropic interactions can help stabilizing complex magnetic structures called multiple-$Q$-states\cite{Kurz2001,Al-Zubi2011,Heinze2011,Kroenlein2018}.
We thus address the potential role of that the CBI might play for such complex magnetic structures, choosing the Fe monolayer on Ir(111) by way of example\cite{Heinze2011}.
The ground state is a nanoskyrmion lattice, which is a type of $2Q$-state made of two symmetry-related wavevectors $\vec{Q}_1$ and $\vec{Q}_2$.
Other combinations lead to further noncollinear states which were calculated to have a similar energy: single-$Q$ spin spirals, the $\vec{Q}_m$-star and $\vec{Q}_m$-vortex states, and the nanovortex lattice.
These magnetic structures are visualized in Fig.~\ref{fig:energy_dependence_structures}a-f, with the triangular N\'eel state included for comparison.
To show how different magnetic interactions affect these states, we write each energy contribution as $E_I = N\,I\,\Gamma_I$, where $N$ is the number of atoms in the magnetic unit cell, $I=\{D,B,C\}$ is the interaction strength, and $\Gamma_I$ is a factor determined by the lattice and the magnetic structure.
Fig.~\ref{fig:energy_dependence_structures}g plots the $\Gamma_I$ factors computed assuming nearest-neighbor pair interactions following $C_{\mathrm{3v}}$ symmetry.
We do not discuss the contribution of $J$, as this interaction is complex and long-ranged for this system\cite{Heinze2011}.
Interestingly, the contribution of the isotropic biquadratic interaction $B$ is similar for every structure, none being particularly favoured.
On the other hand, both the DMI as well as the CBI do favour certain magnetic structures.
The DMI favours almost equally the single-$Q$ spiral and the nanoskyrmion lattice, while the CBI clearly favours the nanoskyrmion lattice ($2Q$ state) over the spin spirals.
This shows that the CBI can play a deciding role in stabilizing noncoplanar multiple-$Q$-states, which have a non-vanishing scalar spin chirality.

\section*{Discussion}
We presented a comprehensive analysis of a new chiral higher-order magnetic pair interaction, the chiral biquadratic interaction (CBI).
Using a microscopic model and a systematic expansion of the electronic grand-potential, we identified the prototypical diagrams behind all kinds of magnetic interactions.
This led us to uncover a new chiral interaction, the CBI, which is linear in the spin-orbit coupling and is the biquadratic equivalent of the DMI, following the same symmetry rules.
In its most general form, this interaction couples four distinct magnetic sites, and consists of terms of the form $\vec{C}_{ijkl}\cdot(\vec{S}_i \times \vec{S}_j)\,(\vec{S}_k \cdot \vec{S}_l)$. We note that a recent study has found signatures of higher-order interactions in magnetic chains on the Re(0001) surface\cite{Laszloffy2019}.
We proved the existence and quantified the importance of this new interaction in relation to other well-known ones by performing systematic first-principles calculations for prototypical systems, magnetic dimers on surfaces with strong spin-orbit coupling.
For these dimers, the CBI is typically around $20-30\%$ of the DMI, reaching $\SI{60}{\percent}$ for a Ni dimer deposited on Pt(111).
Furthermore, we contrasted the properties of the CBI with those of the DMI, revealing that the CBI can simultaneously favor structures with opposite vector spin chiralities (canted ferromagnetic vs.\ canted antiferromagnetic structure), and that the CBI vector is not necessarily parallel to the DMI one.
Insight into the electronic origin of the CBI and the other magnetic interactions was obtained via their dependence on the filling of the electronic states, supporting the proposed mechanism derived from the microscopic model.
We related the magnetic phase diagram of a dimer to that of an infinite chain, and surveyed the implications of the CBI for magnetic monolayers.
Considering several magnetic structures relevant for an Fe monolayer on Ir(111), we showed that the CBI can play a deciding role in favoring a noncoplanar nanoskyrmion lattice over other possible structures.
We thus conclude that the CBI can influence simultaneously the vector and the scalar spin chirality, opening new avenues for stabilizing and engineering complex magnetic textures with both fundamental and technological interest.
We also expect that there should be a non-trivial impact on the related dynamics, transport and topological properties, which should motivate future investigations.


\begin{methods}

\subsection{Density functional theory}
We performed systematic density functional theory calculations with the full-potential Korringa-Kohn-Rostoker (KKR) Green function method\cite{Papanikolaou2002}.
Exchange and correlation effects are treated in the local spin density approximation (LSDA) as parametrized by Vosko, Wilk and Nusair\cite{vosko_accurate_1980}, and SOC is added to the scalar-relativistic approximation\cite{bauer2014}.
The pristine surfaces are modelled by a slab of 22 layers (except for Pt(111) for which 40 layers were used) with the experimental lattice constants with open boundary conditions in the stacking direction, and surrounded by two vacuum regions.
No relaxation of the surface layer of the pristine surface is considered, as it was shown to be negligible\cite{blonski_density-functional_2009}.
We use $150 \times 150$ $k$-points in the two-dimensional Brillouin zone, and the angular momentum expansions for the scattering problem are carried out up to $\ell_\mathrm{max} = 3$.
In the next step, we utilize an embedding method to place each dimer on the fcc-like threefold hollow position of Pt(111) and Ir(111), on the fourfold hollow sites on the Pt(001) surface, and on the hcp-like threefold hollow site of Re(0001).
The embedding region consists of a spherical cluster around each magnetic adatom including nearest-neighbor Pt atoms.
The appropriate structural relaxations of each dimer towards the surface were obtained with the plane-wave code Quantum Espresso\cite{QE-2017}, using ultrasoft scalar relativistic pseudopotentials\cite{pslibrary} with the PBEsol functional\cite{Perdew2008}, and considering a $4\times4$ supercell with $5$ substrate layers (more details and structural information are given in Supplementary Note 4).
All calculations were performed on the supercomputer JURECA at the Forschungszentrum J\"ulich\cite{jureca}.

\subsection{Mapping DFT to an atomistic spin model --- The torque method}
To map the first-principles calculations to an atomistic spin model we adopted the method of constraining fields\cite{Ujfalussy1999}.
The net magnetic moment for each atom in the dimer is defined by averaging the spin magnetization density over its atomic cell, $\vec{m}_i = \int\!\mathrm{d}\vec{r}\;\vec{m}_i(\vec{r}) = |\vec{m}_i|\,\hat{\vec{e}}_i$.
The orientations of the magnetic moments $\hat{\vec{e}}_i$ play the role of emergent degrees of freedom that can be used to define an atomistic spin model,
\begin{equation}
    E_{\mathrm{DFT}}[\{\hat{\vec{e}}\}] = \sum_i\sum_{\alpha,\beta} K_i^{\alpha\beta} \hat{e}_i^\alpha\hat{e}_i^\beta
    + \frac{1}{2}\sum_{i,j}\sum_{\alpha,\beta} J_{ij}^{\alpha\beta} \hat{e}_i^\alpha\hat{e}_j^\beta + \ldots
\end{equation}
To determine the energy of a target magnetic configuration, the total energy functional is augmented by a Zeeman term enforcing the constraint,
\begin{equation}
    E_{\mathrm{cDFT}}[\{\hat{\vec{e}}\}] = E_{\mathrm{DFT}}[\{\hat{\vec{e}}\}] - \sum_i \vec{b}_i \cdot \!\int\!\mathrm{d}\vec{r}\;\vec{m}_i(\vec{r}) \quad.
\end{equation}
The constraining magnetic field is transverse to the orientation of the local magnetic moment, $\vec{b}_i \cdot \hat{\vec{e}}_i = 0$, and it opposes the magnetic force that acts on it if the magnetic structure is not a stationary point of the total energy functional,
\begin{equation}
    \frac{\delta E_{\mathrm{cDFT}}[\{\hat{\vec{e}}\}]}{\delta\hat{\vec{e}}_i} = \frac{\delta E_{\mathrm{DFT}}[\{\hat{\vec{e}}\}]}{\delta\hat{\vec{e}}_i}
    - \vec{b}_i\,|\vec{m}_i| = \vec{0} \quad .
\end{equation}
The induced moments in the surface atoms are allowed to relax without any constraint.
The model parameters are then determined by linear least-squares fitting the constraining fields obtained for a set of self-consistent magnetic configurations to the form of the magnetic force supplied by the atomistic spin model.
The magnetic configurations for the dimers have been chosen using a Lebedev grid\cite{Lebedev1999} containing $14$ directions for each atom, which is well-suited to describe spherical harmonics up to $\ell=2$, resulting in a total of $14^2 = 196$ configurations, which using symmetry arguments (time-reversal invariance of the magnetic energy plus the spatial symmetries that apply on different surfaces) the number of configurations can be further reduced to $56$ for (111) and (0001) surfaces and to $36$ for (001) surfaces.

\end{methods}


\begin{addendum}
 \item This work was supported by the European Research Council (ERC) under the European Union's Horizon 2020 research and innovation program (ERC-consolidator grant 681405 -- DYNASORE). 
 The authors gratefully acknowledge the computing time granted through JARA-HPC on the supercomputer JURECA at the Forschungszentrum J\"ulich\cite{jureca}.
 \item[Author contributions] S.B.\ and M.d.S.D.\ derived the magnetic exchange interactions from the microscopic model and defined the mapping procedure from the first-principles calculations. 
 S.B.\ performed the DFT calculations for all considered systems, the parametrization of the extended magnetic hamiltonian and the model study of several extended magnetic structures.
 All authors analyzed the results and discussed their significance, and collaborated in writing the manuscript.
 \item[Data availability] The authors declare that the main data supporting the findings of this study are available within the article and its Supplementary Information files. 
 Further data is available from the corresponding author upon reasonable request.
 \item[Competing Interests] The authors declare that they have no competing financial interests.
 \item[Correspondence] Correspondence and requests for materials should be addressed to S.B. (email: s.brinker@fz-juelich.de) or to S.L. (email: s.lounis@fz-juelich.de).
\end{addendum}


\bibliography{Lib.bib}

\end{document}